\documentclass[aps,prd,twocolumn,showpacs,preprintnumbers]{revtex4}
\usepackage[dvips]{graphicx}
\usepackage{bm,latexsym,amsmath,amssymb,amsfonts}
\newcommand*{\cP}{{\cal P}}
\begin{document}

\title{
Quantum-mechanical generation of gravitational waves in braneworld
}

\author{Tsutomu~Kobayashi}
\email{tsutomu@tap.scphys.kyoto-u.ac.jp}
\author{Takahiro~Tanaka}
\email{tama@scphys.kyoto-u.ac.jp}

\affiliation{
Department of Physics, Kyoto University, Kyoto 606-8502, Japan 
}


\begin{abstract}
We study the quantum-mechanical generation of 
gravitational waves during inflation on a brane
embedded in a five-dimensional anti-de Sitter bulk.
To make the problem well-posed, we consider the setup in which 
both initial and final phases are given by a de Sitter brane with 
different values of the Hubble expansion rate. 
Assuming that the quantum state is in a de Sitter 
invariant vacuum in the initial de Sitter phase, we 
numerically evaluate the amplitude of quantum fluctuations of 
the growing solution of the zero mode in the final de Sitter phase.  
We find that 
the vacuum fluctuations of the initial Kaluza-Klein gravitons
as well as of the zero mode gravitons contribute to the
final amplitude of the zero mode on small scales,
and the power spectrum is quite well approximated
by what we call the rescaled spectrum, 
which is obtained by rescaling the 
standard four-dimensional calculation following a simple mapping rule.
Our results confirm the speculation raised in Ref.~\cite{Kobayashi:2003cn}
before.

\end{abstract}


\preprint{KUNS-1970}

\maketitle

\section{Introduction}

In recent years
there has been a growing attention
to braneworld scenarios,
in which our observable universe
is realized as a brane embedded in
a higher dimensional bulk spacetime~\cite{Maartens:2003tw}.
Among various models
a simple setup called the Randall-Sundrum type II model~\cite{Randall:1999vf}
is of particular interest
because, despite gravity propagating
into an infinite extra dimension,
four-dimensional general relativity
can be recovered on the brane
due to the warped geometry of anti-de Sitter
spacetime~\cite{Randall:1999vf, Garriga:1999yh, Shiromizu:1999wj}.
Cosmological consequences of the Randall-Sundrum type braneworld
have been widely
discussed~\cite{FRW_brane, Dark_Radiation, Garriga:1999bq, Hawking:2000kj, bulk_inflaton1, bulk_inflaton2},
especially regarding cosmological perturbations~\cite{CP}.

The generation and evolution of perturbations
are among the most important issues in cosmology
because of their direct link to cosmological observations
such as the stochastic gravitational wave background
and the temperature anisotropy of the cosmic microwave
background (CMB),
by which we can probe the early universe.
While the cosmological perturbation theory
in the conventional four-dimensional universe
is rather established~\cite{Kodama:1985bj, Mukhanov:1990me},
calculating cosmological perturbations in the braneworld
still remains to be a difficult problem.
Although some attempts have been made
concerning scalar
perturbations~\cite{Koyama:2002nw, Koyama:2003be, Rhodes:2003ev, Koyama:2004cf, Koyama:2004ap, Yoshiguchi:2004nm},
further progress is awaited
to give a clear prediction about the CMB anisotropy
in the Randall-Sundrum braneworld.
Almost the same is true for gravitational wave (tensor)
perturbations~\cite{Hawking:2000kj, Langlois:2000ns, Gorbunov:2001ge,
Kobayashi:2003cn, Hiramatsu:2003iz, Hiramatsu:2004aa, Ichiki:2003hf,
Ichiki:2004sx, Easther:2003re, Battye:2003ks, Battye:2004qw,
Tanaka:2004ig, Kobayashi:2004wy}. 
However, since their generation and evolution depend basically only on
the background geometry, 
they are slightly easier to handle.

During inflation super-horizon gravitational wave perturbations
are generated from vacuum fluctuations of gravitons. 
The pure de Sitter braneworld is the special case that
allows definite analytical computation of quantum fluctuations. 
Thanks to the symmetry of the de Sitter group,
the perturbation equation becomes separable and 
can be solved exactly~\cite{Langlois:2000ns}.
The time variation of the Hubble parameter generally 
causes mixing of a massless zero mode and
massive Kaluza-Klein modes,
and this effect was investigated 
based on the ``junction'' models,
which assume an instantaneous transition
from a de Sitter to a Minkowski brane~\cite{Gorbunov:2001ge}
or to another de Sitter brane~\cite{Kobayashi:2003cn}.
These works discussed the quantum-mechanical generation of 
gravitational waves within such limited and simplified models.  

As for the classical evolution of gravitational waves
during the radiation dominated epoch, several studies have 
been done~\cite{Hiramatsu:2003iz, Hiramatsu:2004aa, Ichiki:2003hf,
Ichiki:2004sx}, assuming that a given single initial mode for each 
comoving wave number dominates. 
The focus of these works is mainly on 
the evolution of modes which re-enter the horizon 
in the high-energy regime.  
While the late time evolution is worked out analytically 
in Refs.~\cite{Kobayashi:2004wy, Koyama:2004cf}
by resorting to low-energy approximation methods.

In this paper we consider the generation of primordial gravitational
waves
during inflation in more general models of the Randall-Sundrum type.
For definiteness, we adopt a simple setup in which both initial 
and final phases are described by de Sitter braneworlds. 
Two de Sitter phases with different values of the Hubble expansion rate are 
smoothly interpolated. 
This work is an extension of the work on the ``junction'' models
by the present authors in collaboration with 
H.~Kudoh~\cite{Kobayashi:2003cn}. 
In the previous analysis the transition of the Hubble rate was 
abrupt and the gap was assumed to be infinitesimal.  
Here we extend the previous results to more general models 
with smooth transition by
using numerical calculations with a refined formulation.

This paper is organized as follows.
In the next section we briefly summarize past
studies~\cite{Langlois:2000ns, Gorbunov:2001ge, Kobayashi:2003cn} on
the generation of gravitational waves via quantum fluctuations 
during inflation in the Randall-Sundrum braneworld,
emphasizing the mapping formula introduced in 
Ref.~\cite{Kobayashi:2003cn}. 
In Sec.~\ref{formulation} we describe 
our numerical scheme  
to investigate the generation of gravitational waves,
and then in Sec.~\ref{results} we present results of our calculations.
Section~\ref{discussion} is devoted to discussion.


\section{Gravitational waves in inflationary braneworld}\label{past}

\subsection{Pure de Sitter brane}

The background spacetime that we consider is composed of
a five-dimensional AdS bulk, whose metric
is given in the Poincar\'{e} coordinates by
\begin{eqnarray}
ds^2=\frac{\ell^2}{z^2}\left(
-dt^2+\delta_{ij}dx^idx^j+dz^2
\right),\label{Poincare}
\end{eqnarray}
and a Friedmann brane at $z=z(t)$.
Here $\ell$ is the bulk curvature length, which is 
constrained by table-top experiments as 
$\ell \lesssim 0.1$ mm~\cite{Long:2002wn}.

First let us consider the generation of gravitational waves from
pure de Sitter inflation on the brane~\cite{Langlois:2000ns}.
The coordinate system appropriate for the present situation is
\begin{eqnarray}
ds^2=\frac{\ell^2}{\sinh^2 \xi}\left[
\frac{1}{\eta^2}\left(
-d\eta^2+\delta_{ij}dx^idx^j
\right)
+d\xi^2 \right],
\end{eqnarray}
which is obtained from Eq.~(\ref{Poincare})
by a coordinate transformation
\begin{eqnarray}
&&t=\eta\cosh\xi+t_0,
\label{t-dS}\\
&&z=-\eta\sinh\xi,
\label{z-dS}
\end{eqnarray}
where $t_0$ is an arbitrary constant and $\eta$ is the conformal time,  
which is negative.
The de Sitter brane is located at
\begin{eqnarray}
\xi = \xi_b = \mbox{constant},
\end{eqnarray}
and the Hubble parameter on the brane is given by
\begin{eqnarray}
H=\ell^{-1}\sinh\xi_b.
\end{eqnarray}

The gravitational wave perturbations are described by the metric
\begin{eqnarray}
ds^2=\frac{\ell^2}{\sinh^2\xi}\left\{
\frac{1}{\eta^2}\left[
-d\eta^2+(\delta_{ij}+h_{ij})dx^idx^j
\right]
+d\xi^2 \right\},
\end{eqnarray}
and we decompose the perturbations into the spatial Fourier modes as
\begin{eqnarray}
h_{ij}=\frac{\sqrt{2}}{(2\pi M_5)^{3/2}}\int d^3k~
\phi_{\textbf{k}}(\eta, \xi)
e^{i \textbf{k}\cdot\textbf{x}} e_{ij},
\end{eqnarray}
where $e_{ij}$ is the transverse-traceless polarization tensor
and $M_5$ is the fundamental mass scale
which is related to the four-dimensional Planck mass $M_{\rm Pl}$
by $\ell (M_5)^3=M^2_{{\rm Pl}}$.
The prefactor $\sqrt{2}/(M_5)^{3/2}$ is chosen so that
the kinetic term for $\phi_{\textbf{k}}$ in the effective action
is canonically normalized.
Hereafter we will suppress the subscript $\textbf{k}$.
The linearized Einstein equations give the following
Klein-Gordon-type equation for $\phi$:
\begin{eqnarray}
\left[\frac{\partial^2}{\partial \eta^2}
-\frac{2}{\eta}\frac{\partial}{\partial \eta}+k^2
-\frac{\sinh^3\!\xi}{\eta^2}
\frac{\partial}{\partial\xi}
{1\over \sinh^3\!\xi}\frac{\partial}{\partial\xi}\right]\phi=0,
\label{K-G_for_dS}
\end{eqnarray}
which is obviously separable.
Assuming the $Z_2$-symmetry across the brane,
the boundary condition on the brane is given by
\begin{eqnarray}
n^{\mu}\partial_{\mu} \phi|_{{\rm brane}} \propto
\partial_{\xi}\phi|_{\xi=\xi_b}=0,
\end{eqnarray}
where $n^{\mu}$ is an unit normal to the brane.
Equation~(\ref{K-G_for_dS}) admits
one discrete zero mode,
\begin{eqnarray}
\phi_0(\eta) = {\cal N}\cdot\ell^{-1/2}\frac{H}{\sqrt{2k}}
\left(\eta-\frac{i}{k}\right)e^{-ik\eta},
\label{defphi0}
\end{eqnarray}
where ${\cal N}$ is
a normalization constant, which will be fixed later, 
as well as massive Kaluza-Klein (KK) modes,
\begin{eqnarray*}
\phi_{\nu}(\eta, \xi) =\psi_{\nu}(\eta)\chi_{\nu}(\xi),
\end{eqnarray*}
where $\psi_{\nu}(\eta)$ and $\chi_{\nu}(\xi)$ obey
\begin{eqnarray}
&&\left[
\frac{\partial^2}{\partial \eta^2}
-\frac{2}{\eta}\frac{\partial}{\partial \eta}+k^2
+\frac{\nu^2+9/4}{\eta^2}
\right]\psi_{\nu}(\eta)=0,\label{KK-eta}
\\
&&\left[
\sinh^3\!\xi\frac{\partial}{\partial\xi}
{1\over \sinh^3\!\xi}\frac{\partial}{\partial\xi} +\nu^2+\frac{9}{4}
\right]\chi_{\nu}(\xi)=0,\label{KK-xi}
\end{eqnarray}
and the corresponding Kaluza-Klein mass is given by
\begin{eqnarray}
m^2=\left(\nu^2+\frac{9}{4}\right)H^2\geq\frac{9}{4}H^2.
\end{eqnarray}
Due to the mass gap $\Delta m = 3H/2$~\cite{Garriga:1999bq}
between the zero mode and the lowest KK mode,
all the massive KK modes decay on the brane at super-horizon scales during inflation.
The explicit forms of the KK mode functions are shown in Appendix~\ref{KKmodefn}.

In inflationary cosmology, fluctuations in the graviton
field (and other fields such as the inflaton) are
considered to be generated quantum-mechanically.
Following the standard canonical quantization scheme
we quantize the graviton field. 
Treating $\phi$ as an operator, it may be expanded as
\begin{eqnarray}
\phi = \hat a_{0} \phi_0 + \hat a_{0}^{\dagger}\phi_0^*
+\int_0^{\infty}d\nu\left(\hat a_{\nu}\phi_{\nu}
+ \hat a_{\nu}^{\dagger}\phi_{\nu}^* \right).
\end{eqnarray}
The annihilation and creation operators $\hat a_{n}$
and $\hat a_{n}^{\dagger}$ with $n=0$ or $\nu$
satisfy the usual commutation relations
and the modes $\phi_0$ and $\phi_{\nu}$ are normalized so that
they satisfy
\begin{eqnarray}
&&(\phi_0 \cdot \phi_0) = -(\phi_0^*\cdot\phi_0^*) = 1,
\label{Wronskian_condition}\\
&&(\phi_{\nu} \cdot \phi_{\nu'}) = - (\phi_{\nu}^*\cdot\phi_{\nu'}^*)
= \delta(\nu-\nu'),
\nonumber\\
&&(\phi_0 \cdot \phi_{\nu}) = (\phi_{0}^*\cdot\phi_{\nu}^*)
=0,
\nonumber\\
&&(\phi_{n}\cdot\phi_{n'}^*) = 0,
\qquad\mbox{for}\quad n,n'=0, \nu,
\nonumber
\end{eqnarray}
where $(~\cdot~)$ is the Wronskian defined
by~\cite{Gorbunov:2001ge, Kobayashi:2003cn}
\begin{eqnarray}
(X\cdot Y):=-2i\int_{\xi_b}^{\infty}d\xi
\frac{\ell^3}{\eta^2\sinh^3\xi}\left(
X\partial_{\eta}Y^*-Y^*\partial_{\eta}X \right).
\nonumber\\
\end{eqnarray}
From the conditions (\ref{Wronskian_condition}) we can
determine the normalization of the zero mode as
\begin{eqnarray}
{\cal N} = C(\ell H),
\end{eqnarray}
with 
\begin{eqnarray}
C(x):=\left[
\sqrt{1+x^2}+x^2\ln\left(\frac{x}{1+\sqrt{1+x^2}}\right)
\right]^{-1/2}, 
\end{eqnarray}
which is the same as that introduced in
Refs.~\cite{Langlois:2000ns, Gorbunov:2001ge, Kobayashi:2003cn},
and behaves like
$C^2(x)\approx 1$ for $x\ll 1$ and $C^2(x)\approx 3x/2$ for $x\gg 1$.
This function plays an important roll throughout the present paper.

The expectation value of the squared amplitude of the vacuum
fluctuation in the zero mode is given by
\begin{eqnarray}
|\phi_0|^2&=&
\ell^{-1}C^2(\ell H)\frac{H^2}{2k^3}(1+k^2\eta^2)
\nonumber\\
&\to&\ell^{-1}C^2(\ell H)\frac{H^2}{2k^3},
\end{eqnarray}
where the expression in the second line is obtained by
evaluating the perturbation in the super-horizon regime/at a late time,
and hence this is the amplitude of the \textit{growing mode}.
In terms of the power spectrum defined by
\begin{eqnarray}
\cP := \frac{4\pi k^3}{(2\pi)^3}\cdot\frac{2}{(M_5)^3}|\phi_0|^2,
\end{eqnarray}
we have
\begin{eqnarray}
\cP = \frac{2C^2(\ell H)}{M_{\rm Pl}^2}\left(\frac{H}{2\pi}\right)^2,
\label{deSitter_5D}
\end{eqnarray}
where we have used $\ell (M_5)^3=M_{\rm Pl}^2$.
This is the standard flat spectrum up to
the overall factor $C^2(\ell H)$.
In conventional four-dimensional cosmology
the power spectrum of the primordial gravitational waves
from de Sitter inflation is given by~\cite{Liddle:2000cg}
\begin{eqnarray}
\cP_{{\rm 4D}} = \frac{2}{M_{\rm Pl}^2}\left(\frac{H}{2\pi}\right)^2.
\end{eqnarray}
Thus just by rescaling the amplitude in four-dimensional cosmology as
\begin{eqnarray}
H \mapsto HC(\ell H),
\label{mapdS}
\end{eqnarray}
the braneworld result~(\ref{deSitter_5D}) is exactly obtained.

\subsection{The ``junction'' model}

Pure de Sitter inflation on the brane described in the previous subsection
is a special case where the amplitude of the growing zero mode
can be obtained completely analytically.
As a next step to understand gravitational waves
from more general inflation with $H\neq$ const,
a discontinuous change in the Hubble parameter
was considered in Ref.~\cite{Kobayashi:2003cn}. 
In such a ``junction'' model the Hubble parameter is given by
\begin{eqnarray}
H(\eta)=\left\{H_i,\qquad\qquad\qquad \eta<\eta_0,
\atop
H_f=H_i-\delta H,\quad \eta>\eta_0,\right.\label{dis_Hub}
\end{eqnarray}
and
\begin{eqnarray}
\frac{\delta H}{H_i}\ll 1, 
\end{eqnarray}
is assumed. 

The evolution of gravitational wave perturbations
can be analyzed
by solving the equation of motion backward
using the advanced Green's function~\cite{Gorbunov:2001ge}.
A set of mode functions can be constructed
for the initial de Sitter stage,
and another for the final de Sitter stage,
either of which forms a complete orthonormal basis for 
the graviton wave function.
The final (growing) zero mode
may be written as a linear combination of
the initial zero and KK modes.
The Bogoliubov coefficients give the creation rate 
of final zero mode gravitons from the initial vacuum fluctuations
in the Kaluza-Klein modes as well as in the zero mode.
Thus the power spectrum
may be written as a sum of two separate contributions,  
\begin{eqnarray}
\cP = \cP_0+\cP_{{\rm KK}},
\label{sumspectrum}
\end{eqnarray}
where $\cP_0$ and $\cP_{{\rm KK}}$ are the parts 
coming from the initial zero and Kaluza-Klein modes, respectively.

An important result brought by analyzing 
this junction model is that $\cP$ is well approximated by 
the rescaled spectrum $\cP_{{\rm res}}$ 
obtained by using a simple map speculated by 
the exact result of pure de Sitter inflation [Eq.~(\ref{mapdS})].
More precisely,
let $\cP_{{\rm 4D}}(k)$ be
the gravitational wave spectrum evaluated 
in the standard four-dimensional inflationary universe
with the same time dependence of the Hubble parameter~(\ref{dis_Hub}),
and then 
the rescaled spectrum is defined by
\begin{eqnarray}
\cP_{{\rm 4D}}=\frac{2}{M_{{\rm Pl}}^2}
\left(\frac{h_k}{2\pi}\right)^2
\mapsto
\cP_{{\rm res}}=\frac{2C^2(\ell h_k)}{M_{{\rm Pl}}^2}
\left(\frac{h_k}{2\pi}\right)^2.
\label{mapping}
\end{eqnarray}
Comparing the rescaled spectrum with $\cP$ 
obtained from a five-dimensional calculation, one finds that
the difference between these two is suppressed 
to be second order like~\cite{Kobayashi:2003cn}
\begin{eqnarray}
\left|
\frac{\cP-\cP_{{\rm res}}}{\cP}
\right|\lesssim \left(\frac{\delta H}{H_i}\right)^2\ll 1.
\label{agreement}
\end{eqnarray}
It might be worth noting here that
the KK contribution $\cP_{{\rm KK}}$ is necessary
for realizing this interesting agreement
between the braneworld result and the (basically) four-dimensional
result especially at high energies $\ell H\gg 1$.
The above agreement (\ref{agreement}) raised a speculation that 
the mapping formula $h_k\mapsto C(\ell h_k)$
may work with good accuracy
in more general inflation models 
with a smoothly changing expansion rate.
The central purpose of the present paper
is to check whether this speculation is correct or not. 

\vspace{4mm}

\section{Formulation}\label{formulation}

\subsection{Basic equations}

Now let us explain the formulation that we use to
study the generation of gravitational waves
without assuming a pure de Sitter brane.
Our formulation is based on double null coordinates, 
which are presumably the most convenient 
for numerical calculations.

The metric~(\ref{Poincare})
can be rewritten by using double null coordinates
\begin{eqnarray}
&u=t-z,\\
&v=t+z,
\end{eqnarray}
in the form of
\begin{eqnarray}
ds^2=\frac{4\ell^2}{(v-u)^2}\left(
-dudv+\delta_{ij}dx^idx^j
\right).
\end{eqnarray}
The trajectory of the brane can be 
specified arbitrarily by 
\begin{eqnarray}
v=q(u).
\end{eqnarray}
By a further coordinate transformation
\begin{eqnarray}
&&U=u,\\
&&q(V)=v,
\end{eqnarray}
we obtain
\begin{eqnarray}
ds^2=\frac{4\ell^2}{[q(V)-U]^2}\left[
-q'(V)dUdV+\delta_{ij}dx^idx^j
\right],
\end{eqnarray}
where a prime denotes differentiation with respect to the argument. 
Now in the new coordinates the position of the brane is simply given by
\begin{eqnarray}
U=V.
\end{eqnarray}
We will use this coordinate system for actual numerical calculations.

The induced metric on the brane is
\begin{eqnarray}
ds_b^2=\frac{4\ell^2}{[q(V)-V]^2}\left[
-q'(V)dV^2+\delta_{ij}dx^idx^j
\right],
\end{eqnarray}
from which we can read off the conformal time $\eta$ and the scale
factor $a$, respectively, as
\begin{eqnarray}
&&d\eta  = \sqrt{q'(V)} ~dV,\label{eta-V}\\
&&a = \frac{2\ell}{q(V)-V},\label{a-q}
\end{eqnarray}
and hence the Hubble parameter on the brane
is written as
\begin{eqnarray}
\ell H = \frac{1}{2\sqrt{q'(V)}}[1-q'(V)],
\end{eqnarray}
or equivalently
\begin{eqnarray}
q'(V)=\left(\sqrt{1+\ell^2H^2}-\ell H\right)^2.
\label{q-H}
\end{eqnarray}
Given the Hubble parameter as a function of $\eta$,
one can integrate Eqs.~(\ref{eta-V}) and~(\ref{q-H})
to obtain $q$ as a function of $V$.

When the brane undergoes pure de Sitter inflation
(and thus $q'(V)=$ constant),
the following relation between $(U, V)$
and $(\eta, \xi)$ will be useful:
\begin{eqnarray}
&&\xi = \xi_b + \frac{1}{2}\ln\left(\frac{t_0-U}{t_0-V} \right),
\\
&&\eta = -e^{-\xi_b}[(t_0-U)(t_0-V)]^{1/2},
\end{eqnarray}
and
\begin{eqnarray}
q(V) = e^{-2\xi_b}(V-t_0)+t_0.
\label{qVequal}
\end{eqnarray}

The Klein-Gordon-type equation for a gravitational wave perturbation $\phi$
in the $(U, V)$ coordinates
reduces to
\begin{eqnarray}
\!
\left[4\partial_U\partial_V
+\frac{6}{q(V)-U}(\partial_V-q'(V)\,\partial_U)
+q'(V)\,k^2\right]\phi=0,\!\!\!
\cr
\label{KG_in_UV}
\end{eqnarray}
supplemented by the boundary condition
\begin{eqnarray}
[\partial_U-\partial_V]\phi\bigr\vert_{U=V}=0.
\label{bc}
\end{eqnarray}
The expression for the Wronskian evaluated on a constant $V$
hypersurface is given by 
\begin{eqnarray}
(X\cdot Y)={2i}\!\int^{V}_{-\infty}\!\!\!
 dU\! \left[\frac{2\ell}{q(V)-U}\right]^{3}\!
(X\partial_UY^*\!-Y^*\partial_U X),\cr
\end{eqnarray}
which is independent of the choice of the hypersurface.

When inflation on the brane deviates from pure de Sitter one,
the decomposition into the zero mode and KK modes becomes
rather ambiguous.
For this reason 
we require the initial and final phases of inflation to be pure de Sitter,
though arbitrary cosmic expansion is allowed in the intermediate stage.
In both de Sitter phases, $q(V)$ can be fit by Eq.~(\ref{qVequal}), and 
$\xi_b$ and $t_0$ are determined, respectively. 
Hence we have two sets of de Sitter coordinates $(\eta,\xi)$ and 
$(\tilde\eta, \tilde\xi)$. 
We distinguish the coordinates in the final phase by associating 
them with tilde.
In the final de Sitter phase 
the mode will be well outside the horizon,
and hence we expand the graviton field 
in terms of the growing and decaying zero mode solutions $\tilde\phi_g$
and $\tilde\phi_d$ as
\begin{eqnarray}
\phi =\hat A_g\tilde\phi_g+\hat A_d\tilde\phi_d
+
\int d\nu\left(\hat A_{\nu}\tilde\phi_{\nu}
+\hat A_{\nu}^{\dagger}\tilde\phi_{\nu}^*
\right),
\end{eqnarray}
where
\begin{eqnarray}
&&\tilde\phi_g := {\rm Im}\left[\tilde \phi_0\right],\\
&&\tilde\phi_d := {\rm Re}\left[\tilde \phi_0\right],
\end{eqnarray}
and the mode functions with tilde are defined in the same 
way as $\phi_0$ and $\phi_\nu$ with the substitution of 
$(\tilde\xi,\tilde \eta)$ for $(\xi,\eta)$.  
It can be easily seen that the growing and the decaying 
modes are normalized as
\begin{eqnarray}
\left(\tilde\phi_g\cdot \tilde\phi_d\right)_f=\frac{1}{2},\quad
\left(\tilde\phi_g\cdot \tilde\phi_g\right)_f
=\left(\tilde\phi_d\cdot \tilde\phi_d\right)_f=0, 
\label{g-d-Wronski-condi}
\end{eqnarray}
where subscript $f$ means that the expression is to be evaluated in 
the final de Sitter phase. Notice that 
de Sitter mode functions thus defined 
as functions of $(U, V)$ through $(\xi,\eta)$ 
(or $(\tilde\xi,\tilde\eta)$) do not satisfy the equation of 
motion outside the initial (or final) de Sitter phase. 
Back in the initial de Sitter phase the graviton field
can be expanded as
\begin{eqnarray}
\phi=
\hat a_{0} \phi_0 + \hat a_{0}^{\dagger}\phi_0^*
+\int d\nu\left(\hat a_{\nu}\phi_{\nu}
+ \hat a_{\nu}^{\dagger}\phi_{\nu}^* \right).
\label{initial_mode_decompose}
\end{eqnarray}
We assume that initially 
the gravitons are in the de Sitter invariant vacuum state
annihilated by $\hat a_0$ and $\hat a_{\nu}$,
\begin{eqnarray}
\hat a_0|0\rangle = \hat a_{\nu}|0\rangle = 0.
\end{eqnarray}

We would like to evaluate the expectation value of the
squared amplitude of the vacuum fluctuation
in the growing mode at a late time,
\begin{eqnarray*}
\left|\tilde\phi_g\right|^2\langle 0|\hat A_g^2|0\rangle,
\end{eqnarray*}
or equivalently, the power spectrum,
\begin{eqnarray}
\cP (k) &=& \frac{4\pi k^3}{(2\pi)^3}\frac{2}{(M_5)^3}\cdot
\left|\tilde\phi_g\right|^2
\langle 0|\hat A_g^2|0\rangle
\nonumber\\
&\to &\frac{2C^2(\ell H_f)}{M^2_{{\rm Pl}}} \left( \frac{H_f}{2\pi} \right)^2
\langle 0|\hat A_g^2|0\rangle,
\end{eqnarray}
where $H_f$ is the Hubble parameter in the final de Sitter phase.
In order to obtain the final amplitude,
it is not necessary to solve
the evolution of all the (infinite number of) degrees of freedom with their
initial conditions set by Eq.~(\ref{initial_mode_decompose}).
In fact, we have only to 
solve the backward evolution of the final decaying mode,
as explained below.

Suppose that a solution 
$\Phi(U ,V)$ is chosen so as to satisfy 
$\Phi=\tilde\phi_d$ in the final de Sitter phase.
From the Wronskian condition~(\ref{g-d-Wronski-condi})
we see that
\begin{eqnarray*}
{1\over 2} \hat A_g 
&=& \left(\phi\cdot \tilde\phi_d \right)_f
= \left(\phi\cdot \Phi\right)\cr
&=&(\phi_0\cdot \Phi)_i\,\hat a_0+\int d\nu(\phi_{\nu}\cdot \Phi)_i\,
\hat a_{\nu} +\mbox{h.c.}.,
\end{eqnarray*}
where subscript $i$ means that the expression is to be evaluated in 
the initial de Sitter phase.
In the above we used the fact that the Wronskian is constant in time. 
Thus we obtain
\begin{eqnarray}
\langle 0|\hat A_g^2|0\rangle
=4\left[
|(\phi_0\cdot \Phi)_i|^2+\int d\nu|(\phi_{\nu}\cdot \Phi)_i|^2
\right],
\end{eqnarray}
from which we can calculate the power spectrum
of the primordial gravitational waves.
It is obvious that the spectrum is written in the form of 
Eq.~(\ref{sumspectrum}) with 
\begin{eqnarray}
\cP_0&:=& \frac{8 C^2(\ell H_f)}{M^2_{{\rm Pl}}} \left( \frac{H_f}{2\pi} \right)^2
|(\phi_0\cdot \Phi)_i|^2,
\\
\cP_{{\rm KK}}&:=&\frac{8C^2(\ell H_f)}{M^2_{{\rm Pl}}} \left( \frac{H_f}{2\pi} \right)^2
\int d\nu|(\phi_{\nu}\cdot \Phi)_i|^2.
\end{eqnarray}

\subsection{Numerical scheme}

\begin{figure}[t]
  \begin{center}
    \includegraphics[keepaspectratio=true,height=60mm]{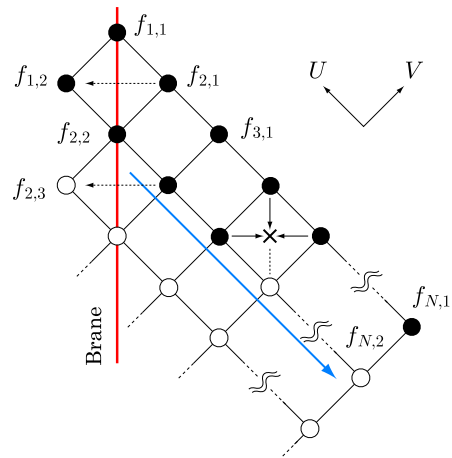}
  \end{center}
  \caption{Numerical (backward) evolution scheme.}
  \label{fig:scheme.eps}
\end{figure}

Here we describe the algorithm that we employ to
solve the backward evolution of the gravitational perturbations numerically.
We set the boundary conditions for the mode $\Phi$ 
so as to be identical to 
a decaying zero mode $\tilde\phi_d$ at a time $V=V_f$ 
in the final de Sitter phase.
We decompose $\Phi$ as 
\begin{eqnarray}
\Phi(U,V) = \tilde\phi_d(U,V) + \delta\Phi(U,V), 
\end{eqnarray}
and solve the equation for $\delta\Phi(U,V)$ instead of 
$\Phi(U,V)$. 
The equation of motion for $\delta\Phi$ is obtained as
\begin{eqnarray}
&&\!\!\!\!\left[4\partial_U\partial_V
+\frac{6}{q(V)-U}(\partial_V-q'(V)\partial_U)
+q'(V)k^2\right]\delta\Phi
\nonumber\\
&&\!\!\!\!=-\left[4\partial_U\partial_V
+\frac{6}{q(V)-U}(\partial_V-q'(V)\,\partial_U)
+q'(V)\,k^2\right]\tilde\phi_d.
\nonumber\\\label{evolution_eq}
\end{eqnarray}
Since both $\Phi$ and $\tilde\phi_d$ satisfy the boundary condition
of the form of Eq.~(\ref{bc}), the boundary condition for $\delta \Phi$
is also written as
\begin{eqnarray}
[\partial_U-\partial_V]\delta\Phi\bigr\vert_{U=V}=0.
\label{bc2}
\end{eqnarray}
We immediately find that we do not have to solve the 
backward evolution in the final de Sitter phase, since  
$\delta\Phi$ identically vanishes there.  

In some cases $\Phi$ is not disturbed so much
from its final configuration $\tilde\phi_d(U,V_f)$, 
especially when we discuss a small deviation from the pure de Sitter 
case. 
It is advantageous then to use a small quantity $\delta \Phi$
as a variable in numerical calculations.
Of course, there is no problem even when 
$\delta \Phi$ does not stay small.

Our scheme to obtain the values of $\delta\Phi$
at the initial surface $V=V_i$ is as follows.
We give the boundary conditions at $V=V_f$ 
as $f_{1,1}, f_{2,1}, \cdots, f_{N,1} =0$,
where 
\begin{eqnarray*}
f_{n,m}:=\delta\Phi(U=V_f-\varepsilon (n-1),V=V_f-\varepsilon (m-1)).
\end{eqnarray*}
A sketch of the numerical grids 
is shown in Fig.~\ref{fig:scheme.eps}. 
Here $N$ is taken to be sufficiently large, 
and we use the same grid spacing $\varepsilon$ 
both in the $U$ and $V$ directions. 
At a virtual site $(U,V)=(V_f, V_f-\varepsilon)$ we set
\begin{eqnarray}
f_{1,2} = f_{2,1},
\end{eqnarray}
so that the boundary condition~(\ref{bc2})
is satisfied at $(V_f-\varepsilon/2, V_f-\varepsilon/2)$.
From $f_{1,1}$, $f_{2,1}$, and $f_{1,2}$ we can determine
the value of $f_{2,2}$ by using the equation of motion~(\ref{evolution_eq})
at $(V_f-\varepsilon/2, V_f-\varepsilon/2)$,
and then from $f_{2,1}$, $f_{3,1}$, and $f_{2,2}$ we can determine $f_{3,2}$,
and so on.
We repeat the same procedure in the subsequent time steps
until we obtain $f_{M,M}, f_{M+1,M}, \cdots, f_{N,M}$
at $V_i=V_f-\varepsilon(M-1)$.

\subsection{A toy model}

\begin{figure}[t]
  \begin{center}
    \includegraphics[keepaspectratio=true,height=76mm]{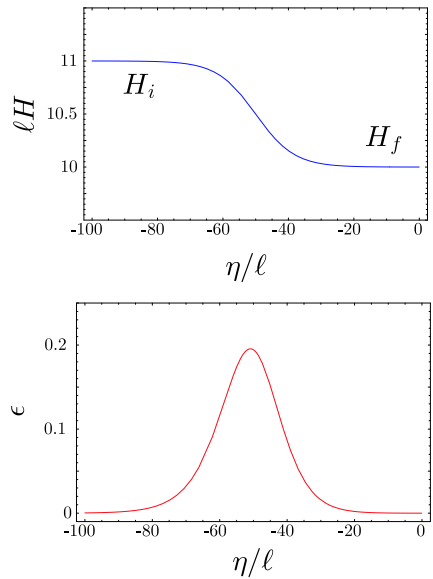}
  \end{center}
  \caption{The Hubble parameter and the slow-roll parameter $\epsilon$
  in our model. This is a plot for the model of Fig.~\ref{fig:high1.eps}.}
  \label{fig:model.eps}
\end{figure}

As stated above, to make the problem well posed, 
we consider models 
which have initial and final de Sitter phases. 
To perform numerical calculations, 
as a concrete example, we adopt a simple toy model 
in which the Hubble parameter is given by
an analytic form
\begin{eqnarray}
H(\eta)=\bar H - \Delta\tanh\left(\frac{\eta-\eta_0}{s}\right),
\end{eqnarray}
where
\begin{eqnarray}
\bar H:=\frac{H_i+H_f}{2},\quad
\Delta :=\frac{H_i-H_f}{2},
\end{eqnarray}
with $H_i$ and $H_f$ 
the initial and final values of the Hubble parameter,
$s$ is a parameter that controls the smoothness of the transition,
and $\eta_0$ indicates a transition time.
Taking the initial time $\eta_i$ and the final time $\eta_f$
so that $\eta_0-\eta_i\gg s$ and $\eta_f-\eta_0\approx -\eta_0\gg s$,
we have $H\approx H_i$ for $\eta \to \eta_i$
and $H\approx H_f$ for $\eta \to\eta_f$.
The scale factor for this inflation model is given by
\begin{eqnarray}
a(\eta)=\left\{
s\Delta \ln\left[2\cosh\left(\frac{\eta-\eta_0}{s}\right)\right]
-\bar H \eta + \Delta \eta_0
\right\}^{-1},
\nonumber\\
\end{eqnarray}
where the integration constant is determined by imposing
that $a\approx (-H_f\eta)^{-1}$ at $\eta \approx \eta_f \approx 0$.
The slow-roll parameter, $\epsilon:=-\partial_{\eta}H/(aH^2)$,
is obtained as
\begin{eqnarray}
\epsilon=
\frac{\Delta\{ s\Delta \ln [2\cosh((\eta-\eta_0)/s)]-\bar H\eta+\Delta \eta_0 \}}
{s[ \bar H \cosh((\eta-\eta_0)/s)-\Delta \sinh((\eta-\eta_0)/s)]^2},
\end{eqnarray}
which takes maximum at $\eta=\eta_0$. The maximum value is
\begin{eqnarray}
\epsilon(\eta_0)=\epsilon_{{\rm max}}
=\frac{\Delta}{s\bar H^2a(\eta_0)}.
\label{ep_max}
\end{eqnarray}
The behavior of the Hubble parameter and the slow-roll parameter
is shown in Fig.~\ref{fig:model.eps}.

Here we should mention the limitation of this simple toy model.
For fixed values of $H_i$, $H_f$, and $\epsilon_{{\rm max}}$
one can prolong the period of the transition $\sim s$ 
measured in conformal time
by taking large $s$. 
However, 
using the definition of $\epsilon_{\rm max}$~(\ref{ep_max}), 
the transition time scale in \textit{proper} time 
is found to be given by 
$\sim a(\eta_0)s
=\bar H^2\epsilon_{\rm max}/\Delta$. 
Therefore we cannot change the transition time scale 
independently of the other parameters, 
$H_i$, $H_f$, and $\epsilon_{{\rm max}}$.

\section{Numerical results}\label{results}

\begin{figure}
  \begin{center}
    \includegraphics[keepaspectratio=true,height=85mm]{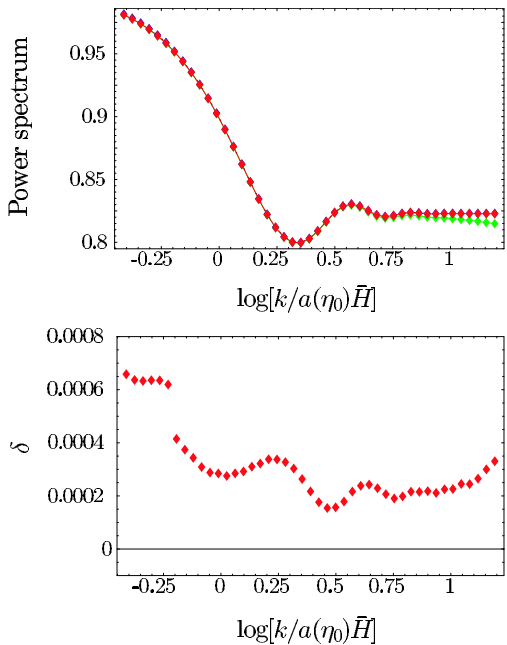}
  \end{center}
  \caption{(color online).
  Top panel: power spectra of gravitational waves
  normalized by $(2C^2(\ell H_i)/M_{{\rm Pl}}^2)(H_i/2\pi)^2$.
  Red diamonds (upper ones) indicate the result including the
  contributions from the initial Kaluza-Klein modes,
  while green diamonds (lower ones) 
  represent the contribution from the initial zero mode.
  Although the rescaled four-dimensional spectrum
  is shown by blue diamonds, they are almost hidden by
  the red ones since the result of the five-dimensional
  calculation is well approximated by the rescaled four-dimensional spectrum.
  Bottom panel: difference between the five-dimensional and
  rescaled four-dimensional spectra.}%
  \label{fig:low1.eps}
\end{figure}
\begin{figure}
  \begin{center}
    \includegraphics[keepaspectratio=true,height=85mm]{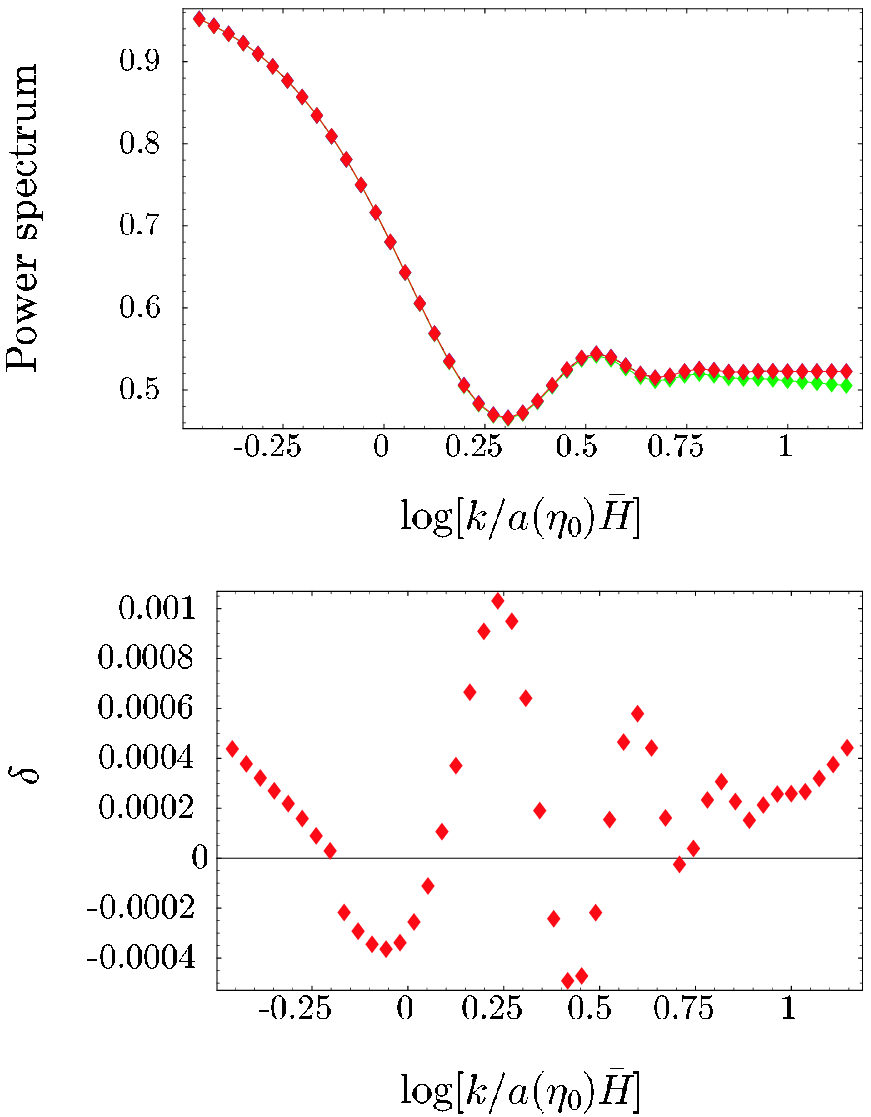}
  \end{center}
  \caption{Same as Fig.~\ref{fig:low1.eps}, but the parameters are different.}%
  \label{fig:low2.eps}
\end{figure}
\begin{figure}
  \begin{center}
    \includegraphics[keepaspectratio=true,height=85mm]{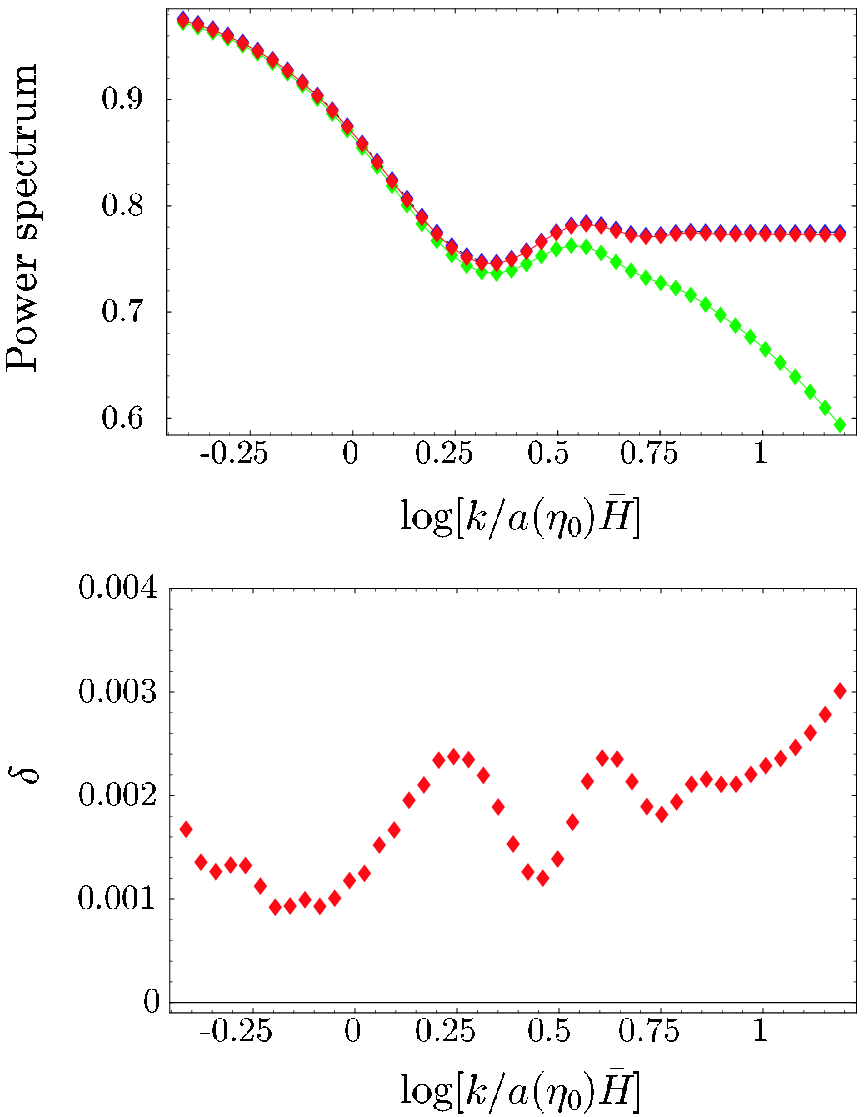}
  \end{center}
  \caption{Same as Fig.~\ref{fig:low1.eps}, but the parameters are different.}%
  \label{fig:middle1.eps}
\end{figure}
\begin{figure}
  \begin{center}
    \includegraphics[keepaspectratio=true,height=85mm]{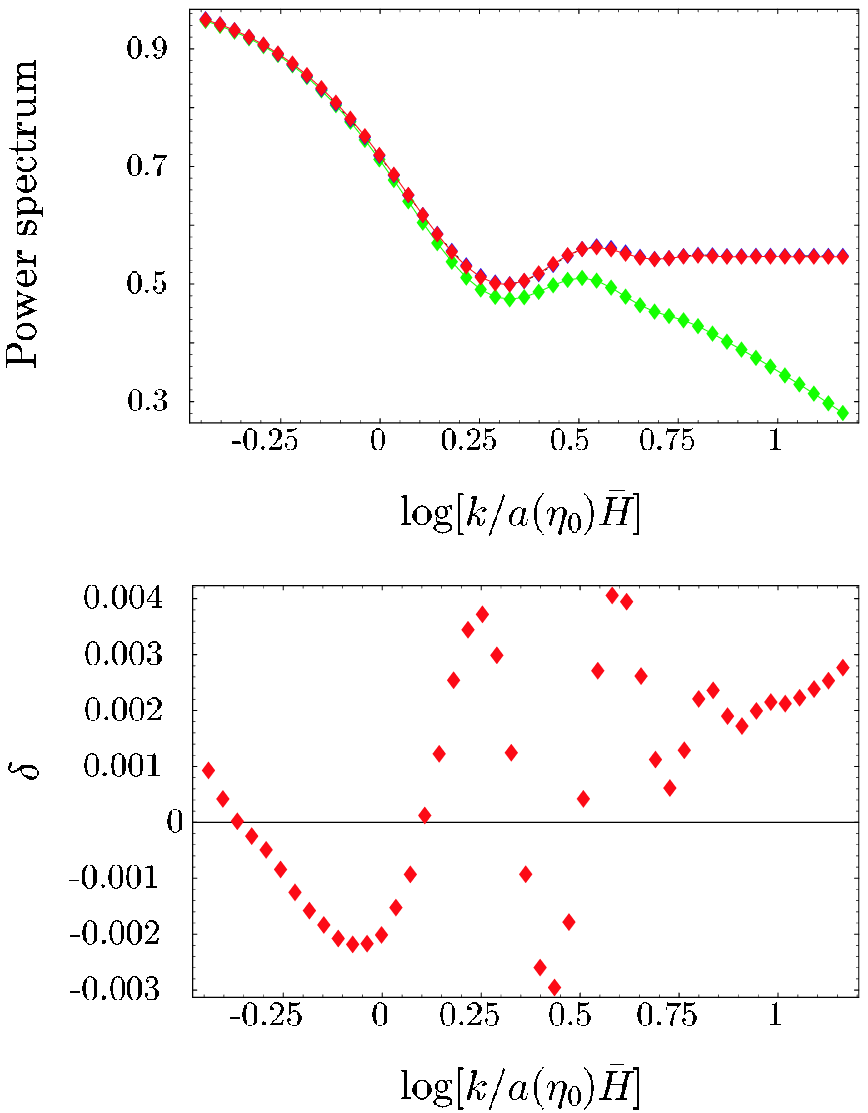}
  \end{center}
  \caption{Same as Fig.~\ref{fig:low1.eps}, but the parameters are different.}%
  \label{fig:middle2.eps}
\end{figure}
\begin{figure}
  \begin{center}
    \includegraphics[keepaspectratio=true,height=85mm]{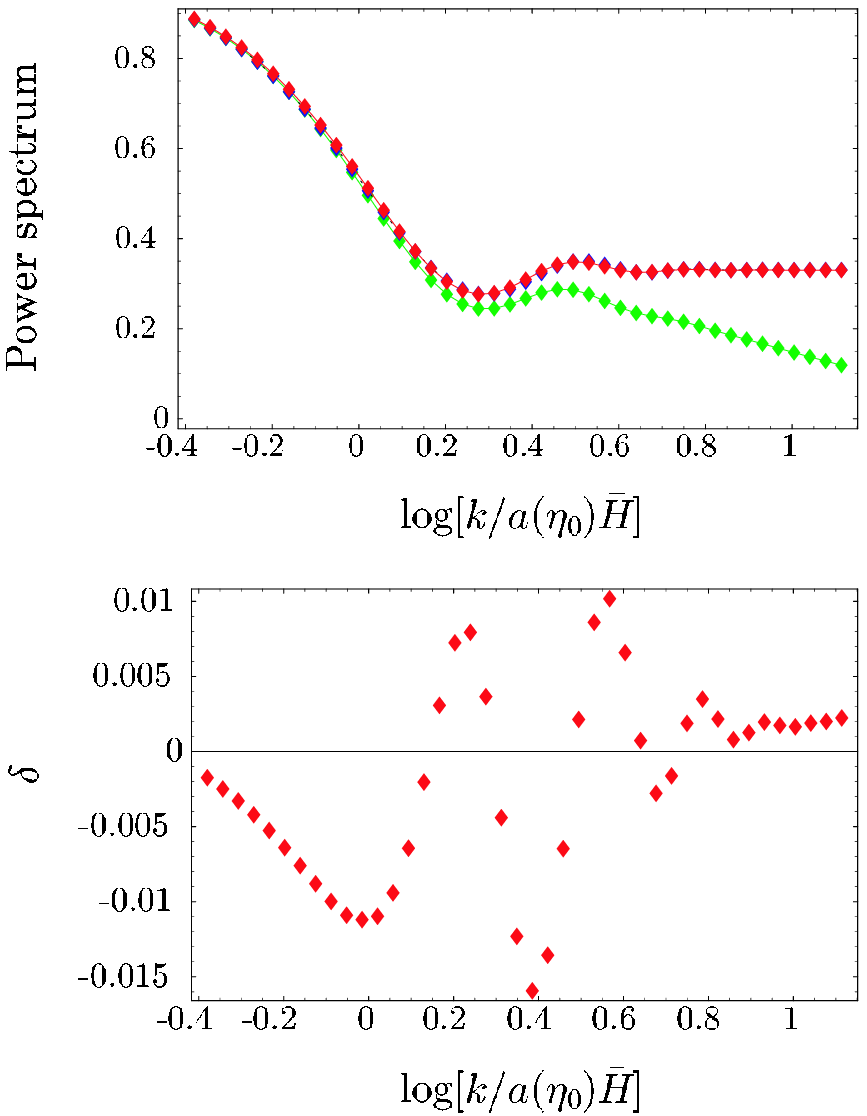}
  \end{center}
  \caption{Same as Fig.~\ref{fig:low1.eps}, but the parameters are different.}%
  \label{fig:middle3.eps}
\end{figure}
\begin{figure}
  \begin{center}
    \includegraphics[keepaspectratio=true,height=85mm]{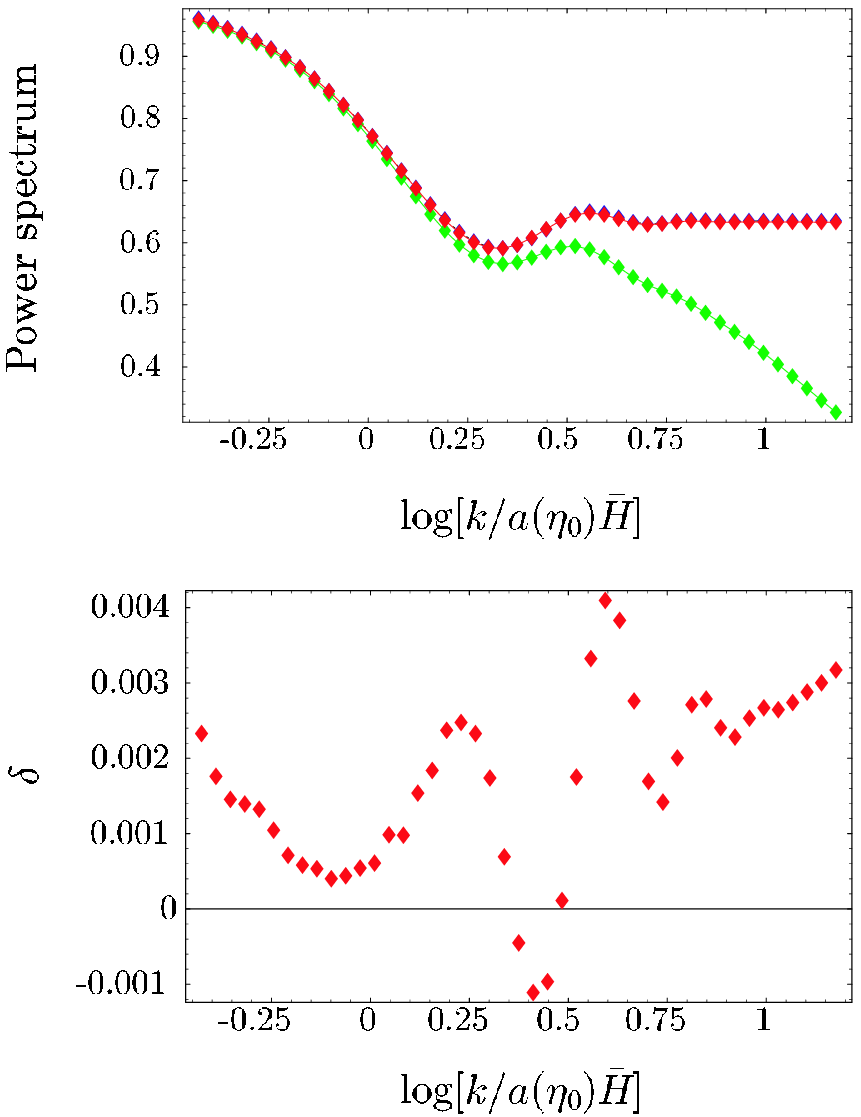}
  \end{center}
  \caption{Same as Fig.~\ref{fig:low1.eps}, but the parameters are different.}%
  \label{fig:middle-high.eps}
\end{figure}
\begin{figure}
  \begin{center}
    \includegraphics[keepaspectratio=true,height=85mm]{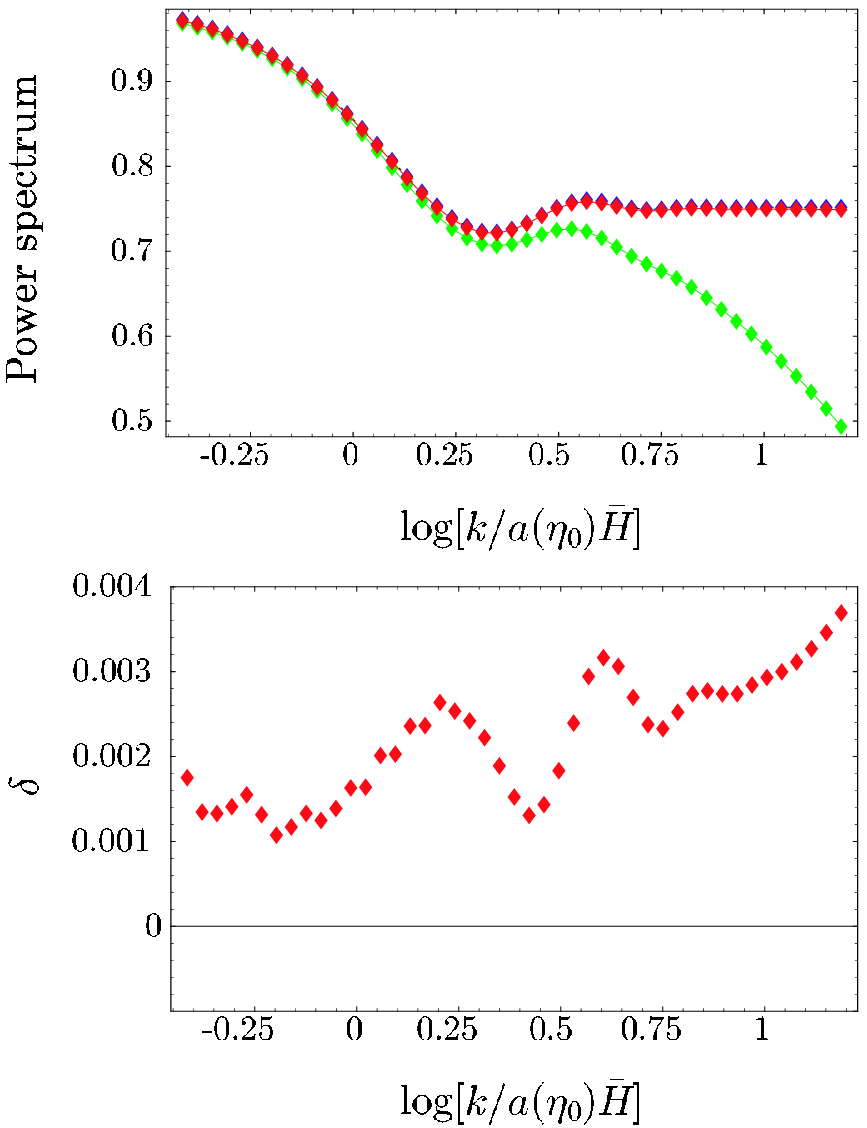}
  \end{center}
  \caption{Same as Fig.~\ref{fig:low1.eps}, but the parameters are different.}%
  \label{fig:high1.eps}
\end{figure}
\begin{figure}
  \begin{center}
    \includegraphics[keepaspectratio=true,height=85mm]{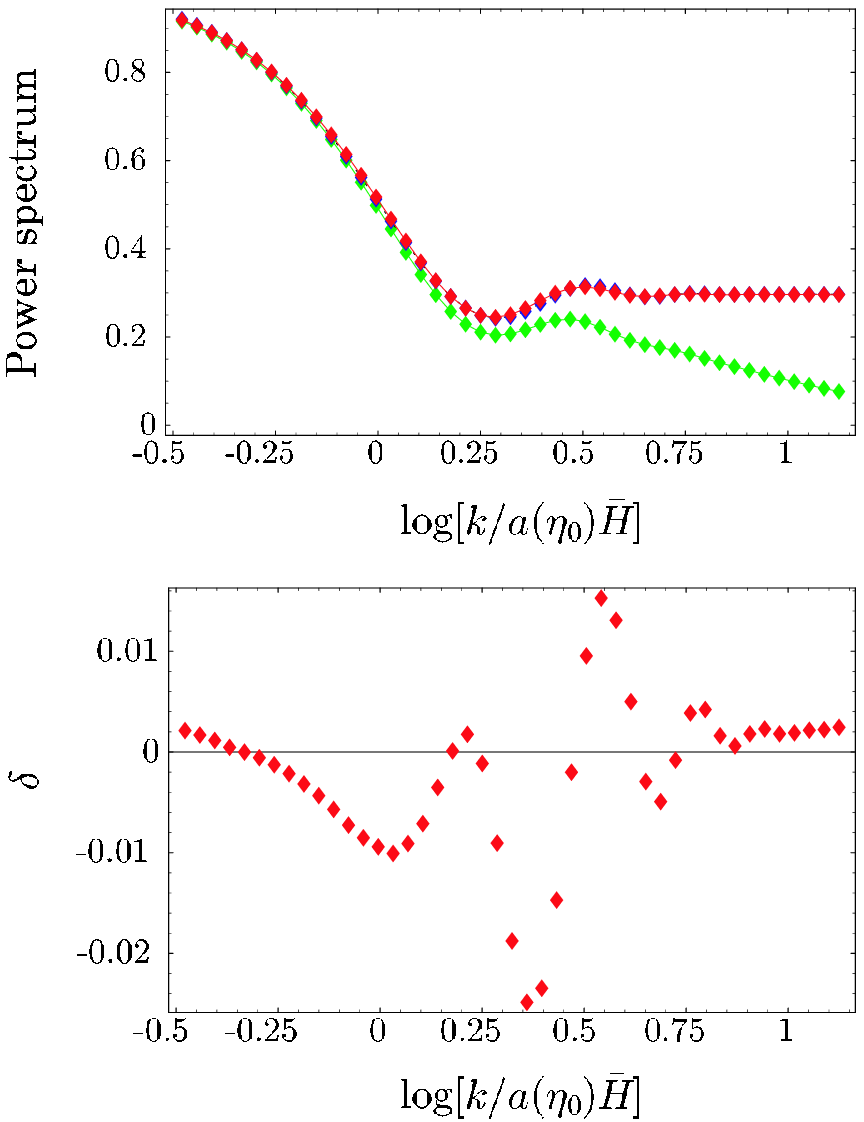}
  \end{center}
  \caption{Same as Fig.~\ref{fig:low1.eps}, but the parameters are different.}%
  \label{fig:high2.eps}
\end{figure}

\begin{figure}
  \begin{center}
    \includegraphics[keepaspectratio=true,height=35mm]{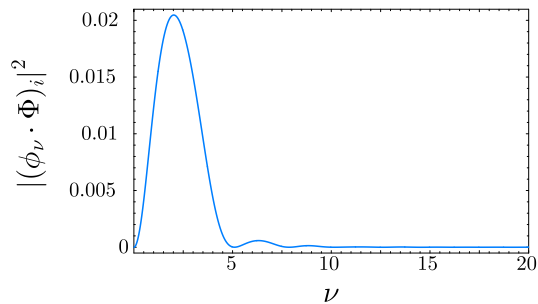}
  \end{center}
  \caption{Wronskian $|(\phi_{\nu}\cdot\Phi)_i|^2$ as a function of $\nu$.
  The parameters are given by those of Fig.~\ref{fig:high1.eps} and
  $k/a(\eta_0)\bar H=10$.}%
  \label{fig:KK.eps}
\end{figure}

\begin{table}
\caption{\label{tab:table1}
Parameters used for the numerical calculations
presented in the figures.
}
\begin{ruledtabular}
\begin{tabular}{l|rrr|rr}
&$\ell H_i$&$\ell H_f$&$\epsilon_{{\rm max}}$&$\ell/s$&$\eta_0/\ell$\\
\hline
Figure~\ref{fig:low1.eps}  & 0.11 & 0.10 & 0.21 & 0.09 & $-50.45$\\
Figure~\ref{fig:low2.eps}  & 0.11 & 0.08 & 0.69 & 0.1 & $-50.46$\\
Figure~\ref{fig:middle1.eps}  & 1.1 & 1.0 & 0.21 & 0.09 & $-50.21$\\
Figure~\ref{fig:middle2.eps}  & 1.25 & 1.0 & 0.45 & 0.09 & $-50.21$\\
Figure~\ref{fig:middle3.eps}  & 1.5 & 1.0 & 0.75 & 0.09 & $-50.21$\\
Figure~\ref{fig:middle-high.eps}  & 3.5 & 3.0 & 0.34 & 0.095 & $-50.08$\\
Figure~\ref{fig:high1.eps}  & 11 & 10 & 0.19 & 0.085 & $-50.02$\\
Figure~\ref{fig:high2.eps}  & 15 & 10 & 0.75 & 0.09 & $-50.02$\\
\end{tabular}
\end{ruledtabular}
\end{table}

Using various parameters shown in Table.~\ref{tab:table1}
we performed numerical calculations,
the results of which are presented in
Figs.~\ref{fig:low1.eps}--\ref{fig:high2.eps}. 
Numbers of grids are $N\times M=50000 \times 1000$, and
the grid separation $\varepsilon$ is chosen to be
$(0.11 - 2.5)\times \ell$
depending on the energy scale of
inflation.
Note here that the step width in conformal time $\eta$
is given by $\Delta\eta=\sqrt{q'(V)}~\varepsilon$ [Eq.~(\ref{eta-V})],
and $q'(V)$ becomes smaller for a larger value of $\ell H$ [Eq.~(\ref{q-H})].
Our choice of $\varepsilon$ makes $\Delta\eta$
about the same size, $\Delta\eta\approx 0.1$,
in all the calculations.
Integration over $\nu$ is performed up to $\nu=20$ -- $40$ 
with an equal grid spacing of $0.15$.
The typical behavior of the integrand $|(\phi_{\nu}\cdot \Phi)_i|^2$
is shown in Fig.~\ref{fig:KK.eps}.

In order to compare the five-dimensional power spectrum
with a four-dimensional counterpart,
we solve the conventional evolution equation
for gravitational waves in four-dimensions,
\begin{eqnarray}
\left(\partial_{\eta}^2+2aH\partial_{\eta} +k^2\right)\Phi =0,
\end{eqnarray}
where $H(\eta)$ is given by the same function
as used for the corresponding five-dimensional computation,
and then we calculate the \textit{rescaled} power spectrum 
obtained from the four-dimensional `bare' spectrum
by using the mapping formula (\ref{mapping}).

We introduce a parameter that represents
a difference between the five-dimensional 
power spectrum and the rescaled four-dimensional spectrum:
\begin{eqnarray}
\delta(k) := \frac{\cP_{{\rm res}} - \cP}{\cP}.
\end{eqnarray}
In all the cases of our calculations we clearly see that
$|\delta| \ll 1$ and thus we conclude that
\textit{the primordial spectrum of the gravitational waves
in the braneworld
is quite well approximated by the rescaled four-dimensional spectrum}.
We also find that the difference becomes larger for larger values of
the slow-roll parameter $\epsilon$,
but $|\delta|$ is no greater than ${\cal O}(10^{-2})$
even for models with $\epsilon_{{\rm max}}\simeq 0.75$.
Notice that the universe is not inflating any more if $\epsilon$ is 
greater than unity. 
At high energies $\ell H \gtrsim 1$
it can be seen that dependence of $\delta$ on $\ell H$ 
is weak, while at low energies like $\ell H \sim 0.1$,
$\delta$ is further suppressed compared to the high energy cases.

Contributions from the initial KK modes can be significant
for small wavelength modes.
For example, for the modes with $k/a(\eta_0)\bar H \simeq 10$
we see that $\cP_{{\rm KK}}/\cP\sim{\cal O}(10^{-2})$
at low energies,
but the KK contribution can become as large as $\cP_{{\rm KK}}/\cP\sim 0.6$
in our most `violent' model with $\ell H\sim 10$ and $\epsilon_{{\rm max}}\simeq 0.75$.

\section{Discussion}\label{discussion}

In this paper we have investigated
the generation of gravitational waves during inflation on a brane
and computed the primordial spectrum.
Extending the previous work~\cite{Kobayashi:2003cn}, 
we have considered a model composed of the initial
and final de Sitter stages, and the transition region
connecting them smoothly.
We have numerically solved the backward evolution
of the final \textit{decaying} mode
to obtain the amplitude of the \textit{growing} zero mode
at a late time
by making use of the Wronskian method,

We found that the power spectrum $\cP$
is well approximated by the rescaled spectrum $\cP_{{\rm res}}$
basically calculated in standard four-dimensional inflationary cosmology.
Here the rescaling formula for the amplitude
is given by a simple map $h_k \mapsto h_kC(\ell h_k)$,
where $C$ is the normalization factor of the zero mode.
Although the difference of the two spectra,
$\delta:=(\cP_{{\rm res}}-\cP)/\cP$,
depends on the energy scale of inflation and the slow-roll parameter,
our numerical analysis clearly shows that
in any case the mapping formula works with quite good accuracy,
yielding $|\delta| \lesssim {\cal O}(10^{-2})$.
This implies that the mapping relation
between the two spectra holds
quite generally
in the Randall-Sundrum braneworld,
which we believe is a useful formula.

We have taken into account the
vacuum fluctuations in initial Kaluza-Klein modes as well as the zero mode,
and found that both of them contribute 
to the final amplitude of the zero mode.
The amount of the initial KK contribution
can be large on small scales,
and it increases as the energy scale of inflation $\ell H$ becomes 
higher.
This gives rise to a quite interesting picture.
When the expansion rate changes during inflation,
zero mode gravitons escape into the bulk as KK gravitons,
but at the same time bulk gravitons
come onto the brane to compensate for the loss,
and these two effects almost cancel each other.
This seems to happen,
irrespective of the energy scale,
in a wide class of the inflation models
even with a not-so-small slow-roll parameter.
It is suggested that this is not the case in the radiation dominated
decelerating
universe~\cite{Hiramatsu:2003iz, Hiramatsu:2004aa, Ichiki:2003hf, Ichiki:2004sx},
where the decay into KK gravitons reduces the amplitude
of the gravitational waves on the brane.
The junction model of Ref.~\cite{Gorbunov:2001ge}
joining de Sitter and Minkowski branes
also show the suppression of the gravitational wave amplitude
at high energies.

What is the reason for the remarkable agreement
of the braneworld spectrum and the rescaled four-dimensional spectrum?
Extremely long wavelength modes,
which leave the horizon
during the initial de Sitter stage much before
the Hubble parameter changes,
have a squared
amplitude of $(2C^2(\ell H_i)/M^2_{{\rm Pl}})(H_i/2\pi)^2$, 
and the amplitude of the perturbations stays constant
during the subsequent stages~\footnote{In the $k^2\to 0$ limit,
$\phi$=const is a growing solution of Eq.~(\ref{KG_in_UV}).}.
The same is true for four-dimensional inflationary cosmology, and so
the mapping formula is applicable to these long wavelength modes.
On the other hand, a mode whose wavelength
is much shorter than the Hubble horizon scale at the transition time
will feel the transition as adiabatic and hence the particle production
is exponentially suppressed,
$
\langle 0|\hat A_g^{\dagger} \hat A_g |0\rangle \approx 1,
$
leading to $\cP\approx (2C^2(\ell H_f)/M^2_{{\rm Pl}})(H_f/2\pi)^2$.
The same argument can be applied to the conventional cosmology.
Thus it is not surprising that the mapping formula works
for such short wavelength modes.
However, at present there seems no simple reason
for the amplitude of the modes with $k\sim {\cal O}(a(\eta_0)\bar H)$ to
coincide with the rescaled one.

\acknowledgments
We thank Shinji Mukohyama for valuable discussions. 
This work is supported in part by JSPS Research Fellowships for
Young Scientists No.~1642, 
by Grant-in-Aid for Scientific Research, Nos.
14047212 and 16740141, 
and by that for the 21st Century COE
``Center for Diversity and Universality in Physics'' at 
Kyoto university, both from the Ministry of
Education, Culture, Sports, Science and Technology of Japan.

\appendix

\section{Kaluza-Klein mode functions}\label{KKmodefn}

In the de Sitter braneworld the Kaluza-Klein mode functions
in the time direction satisfy Eq.~(\ref{KK-eta}) and are given by
\begin{eqnarray}
\psi_{\nu}(\eta) = \frac{\sqrt{\pi}}{2}\ell^{-3/2}
e^{-\pi\nu/2}(-\eta)^{3/2}H^{(1)}_{i\nu}(-k\eta),
\end{eqnarray}
where $H^{(1)}_{i\nu}$ is the Hunkel function.
They are normalized so that
\begin{eqnarray}
-\frac{i\ell^3}{\eta^2}\left(
\psi_{\nu}\partial_{\eta}\psi^*_{\nu}-\psi_{\nu}^*\partial_{\eta}\psi_{\nu}
\right) = 1.
\end{eqnarray}
The mode functions in the extra direction satisfies Eq.~(\ref{KK-xi})
and are given by
\begin{eqnarray}
\chi_{\nu}(\xi) &=& C_1(\sinh\xi)^2
\left[
P^{-2}_{-1/2+i\nu}(\cosh\xi)\right.
\nonumber\\
&&\quad\left.-C_2
Q^{-2}_{-1/2+i\nu}(\cosh\xi) \right],
\end{eqnarray}
where $P^{-2}_{-1/2+i\nu}$ and $Q^{-2}_{-1/2+i\nu}$
are the associated Legendre functions.
The constants $C_1$ and $C_2$ are determined by
the normalization condition
\begin{eqnarray}
2\int^{\infty}_{\xi_b}\frac{d\xi}{(\sinh\xi)^3}\chi^*_{\nu'}\chi_{\nu}
=\delta(\nu-\nu'),
\end{eqnarray}
and the boundary condition~(\ref{bc}), respectively, as
\begin{eqnarray}
C_1&=&\left[\left| \frac{\Gamma(i\nu)}{\Gamma(5/2+i\nu)}\right|^2\right.
\nonumber\\
&&\hspace{-10mm}
\left.+\left| \frac{\Gamma(-i\nu)}{\Gamma(5/2-i\nu)}
-\pi C_2\frac{\Gamma(i\nu-3/2)}{\Gamma(1+i\nu)}\right|^2\right]^{-1/2},
\\
C_2&=&
\frac{P^{-1}_{-1/2+i\nu}(\cosh\xi_b)}{Q^{-1}_{-1/2+i\nu}(\cosh\xi_b)}.
\end{eqnarray}




\end{document}